\documentclass
[twocolumn,aps,prc,amsmath,amssymb,floatfix,nofootinbib]
{revtex4-1}
\usepackage{CJK}                     
\usepackage[dvips]{graphicx}
\usepackage{mathptmx}                
\usepackage{bm}                      
\usepackage{array}
\usepackage{multirow}                
\usepackage{xcolor}                  %

\def\bea{\begin{eqnarray}} \def\eea{\end{eqnarray}}
\def\beq{\begin{equation}} \def\eeq{\end{equation}}
\def\bal#1\eal{\begin{align}#1\end{align}}
\def\bse#1\ese{\begin{subequations}#1\end{subequations}}
 
\def\rv{\bm{r}}
\def\kv{\bm{k}}
\def\Kv{\bm{K}}
\def\tv{\bm{\tau}}
\def\sv{\bm{\sigma}}
\def\tt{(\tv_1\cdot\tv_2)}
\def\ss{(\sv_1\cdot\sv_2)}
\def\eps{\epsilon}
\def\om{\omega}
\def\be{\beta}
\def\fm3{\;\text{fm}^{-3}}
\def\mev{\;\text{MeV}}

\begin{document}

\title{
Nucleon effective mass in hot dense matter}
\author{
X. L. Shang,$^{1,2}$
A. Li,$^{3}$
Z. Q. Miao,$^{3}$
G. F. Burgio,$^{4}$
H.-J. Schulze$^{4}$}
\affiliation{
$^1$
Institute of Modern Physics, Chinese Academy of Sciences, Lanzhou 730000, China\\
$^2$
School of Nuclear Science and Technology,
University of Chinese Academy of Sciences, Beijing 100049, China\\
$^3$
Department of Astronomy, Xiamen University, Xiamen, Fujian 361005, China\\
$^4$
Sezione INFN, Dipartimento di Fisica, Universit\`a di Catania,
Via Santa Sofia 64, I-95123 Catania, Italy
}
\date{\today}

\begin{abstract}
Nucleon effective masses are studied in the framework of the
Brueckner-Hartree-Fock many-body approach at finite temperature.
Self-consistent calculations using the Argonne $V_{18}$ interaction
including microscopic three-body forces are reported
for varying temperature and proton fraction up to several times
the nuclear saturation density.
Our calculations are based on the exact treatment of the center-of-mass momentum
instead of the average-momentum approximation employed in previous works.
We discuss in detail the effects of the temperature together with those of the
three-body forces, the density, and the isospin asymmetry.
We also provide an analytical fit of the effective mass
taking these dependencies into account.
The temperature effects on the cooling of neutron stars are briefly discussed
based on the results for betastable matter.
\end{abstract}


\maketitle

\section{Introduction}

The nucleon effective mass and its dependence on density and temperature,
$m^*(\rho,T)$,
serve as important microscopic input for the study of the thermal properties
(e.g., thermal conductivity, specific heat, neutrino reaction rates)
of (proto) neutron stars (NSs)
\cite{Pag04,Bal12,Tar16,Deh16,Deh18,For18,Sht18,Sht18plb,Lib18,%
Pot18,Pot19,Wei19}.
For cold dense matter,
microscopic nuclear many-body calculations have been performed, for example,
starting from a realistic two-body potential
plus a three-body force (TBF)
within the Brueckner-Hartree-Fock (BHF) formalism
\cite{Bal88,Zuo05,Bal14,Li16,Bal17},
and within the Dirac-Brueckner-Hartree-Fock (DBHF) method \cite{Haa87,Sam10}.
The calculations have been done up to around $5\rho_0$,
for both asymmetric nuclear matter and beta-stable NS matter,
with $\rho_0=0.17\fm3$ being the nuclear saturation density.
The dependence of the nucleon effective mass on both density
$\rho=\rho_n+\rho_p$ and isospin asymmetry
$\be=(\rho_n-\rho_p)/\rho$,
where $\rho_n$ and $\rho_p$ are the neutron and proton number densities,
has been included in fitting formulas \cite{Bal14}
for easy implementation in astrophysical applications.

Thermal effects are known to be important
\cite{Bom94,Pra97,Oer17,Fio18,Rai19}
for the study of proto neutron stars (PNSs),
core collapse supernovae,
binary NS mergers, black-hole accretion disks, etc.
There are several attempts to construct a finite-temperature
equation of state (EOS),
based on a Skyrme nuclear force \cite{Lat91},
on relativistic mean field theory \cite{She99,Hem10},
or within microscopic models
\cite{Fri81,Lej86,Hub98,Rio05,Nic06,Pen08,Muk09,BS10,Li10,BSL11,%
Che12,Li13,Li15,Car19n,Car19s,Lu19}.
The purpose of this paper is to report a systematic study of the nucleon
single-particle (s.p.) properties
on a microscopic basis for hot nuclear/NS matter.
We will concentrate on the neutron/proton effective mass
with varying temperature and proton fraction,
for broad use in these dynamical phenomena.

For this purpose, we employ the BHF model \cite{bbg1,bbg2}
extended to asymmetric nuclear matter and finite temperature \cite{bbg3,bbg4}.
The realistic Argonne $V_{18}$ two-body nucleon-nucleon ($NN$) potential
\cite{av18} is used,
together with the consistent microscopic TBF \cite{tbf1,tbf2,tbf3,tbf4} for
correctly reproducing the empirical saturation point of symmetric nuclear matter.
Previously, the temperature dependence of the effective mass has been studied
within BHF with or without the inclusion of TBF
\cite{bbg4,Lej86,Bal88,Bom94,Gra87,Bom06,Zuo06}.
In the present study,
we use the exact expression of the angular integration
for the center-of-mass (c.m.) momentum
to improve the reliability and the convergence of the BHF code.
In earlier BHF studies an average-c.m.-momentum approximation
was usually adopted,
which could lead to different predictions for high-order contributions
in describing the bulk properties for nuclear matter and the EOS \cite{th1},
and should be improved in the studies of nucleon s.p.~properties.

The paper is organized as follows.
We provide the BHF formalism for hot asymmetric nuclear matter in Sec.~II,
including the extension to full evaluation of the c.m.~momentum.
Sec.~III presents
the s.p.~effective masses in both nuclear matter and NS matter,
together with their analytic fitting formula.
Sec.~IV gives a summary of this work.

\section{Formalism}
\label{s:form}

\subsection{Effective masses in the BHF approach}

The calculations for hot asymmetric nuclear matter are based on the
Brueckner-Bethe-Goldstone (BBG) theory \cite{bhf1,bhf2,bhf3,bbg1,bbg2}
and the extension to finite temperature \cite{Lej86,bf99,bbg3,bbg4}.
Here we simply give a brief review for completeness.
The starting point in Brueckner theory is the effective reaction matrix $G$,
which satisfies the generalized Bethe-Goldstone (BG) equation ($\tau=n,p$),
\bal
 & \langle 12|G_{\tau\tau'}(\om,T)|1'2'\rangle =
 \langle 12|V_{\tau\tau'}|1'2'\rangle
 + \sum_{1''2''} \langle12|V_{\tau\tau'}|1''2''\rangle
\nonumber\\&\times
 \frac{Q_{\tau\tau'}}{\om-e_{\tau}(1'')-e_{\tau'}(2'')}
 \langle 1''2''|G_{\tau\tau'}(\om,T)|1'2'\rangle \:,
\label{e:bge}
\eal
where $\om$ is the so-called starting energy,
$V=V_{NN}+V_3^{\rm eff}$ is the employed
Argonne $V_{18}$ $NN$ interaction \cite{av18}
plus an effective two-body force derived from a microscopic TBF
\cite{tbf1,tbf2,tbf3,tbf4},
and $1\equiv(\kv_1,\sigma_1)$
etc.~denote the momentum and spin $z$ components.
For non spin-polarized nuclear matter,
the spin-up and spin-down states are degenerate
and hereafter we omit the spin index.
The Pauli operator at finite temperature reads
\beq
 Q_{\tau\tau'} = Q_{\tau\tau'}(\kv_1,\kv_2,T) =
 \big[1-f_\tau(\kv_1,T)\big] \big[1-f_{\tau'}(\kv_2,T)\big]
\eeq
with the Fermi distribution
\beq
 f_\tau(\kv,T) =
 \bigg[ 1 + \exp\Big(\frac{e_\tau(\kv)-\tilde{\mu}_\tau}{T}\Big) \bigg]^{-1}
\:.
\eeq
The auxiliary chemical potential $\tilde{\mu}_\tau$
can be calculated from the following
implicit equation for any fixed density and temperature~\cite{Lej86}:
\beq
 \rho_\tau = \sum_{\kv} f_\tau(\kv,T) \:.
\eeq
In BHF approximation, the s.p.~energy is given by
\beq
 e_\tau(\kv) \equiv e_\tau(\kv,T)
 = \frac{\kv^2}{2m} + U_\tau(\kv,T) \:,
\eeq
where the s.p.~potential $U_\tau(\kv)$
is obtained from the real part of the on-shell antisymmetrized $G$ matrix, i.e.,
\beq
 U_\tau(\kv) = \sum_{\kv'\tau'}f_{\tau'}(\kv',T)\,
 \text{Re} \langle\kv\kv' |
 G_{\tau\tau'}[e_\tau(\kv)+e_{\tau'}(\kv'),T] |
 \kv\kv'\rangle_{A} \:.
\label{e:u}
\eeq
Eqs.~(\ref{e:bge},4,5,6) are then solved self-consistently
for given density $\rho$, isospin asymmetry $\be$, and temperature $T$.
The $G$ matrix, the auxiliary chemical potentials $\tilde{\mu}_\tau$,
and the s.p.~potential $U_{\tau}(k)$
are all implicitly dependent on $\rho$, $\be$, and $T$.
Regarding the physical observables we will study here,
the effective mass $m_\tau^*$ can be calculated from the s.p.~energy as
\beq
 \frac{m_\tau^*(k)}{m} =
 \frac{k}{m}\Big[\frac{de_\tau(k)}{dk}\Big]^{-1} \:,
\label{e:ms}
\eeq
where $m$ is the bare nucleon mass.
It depends on $\rho$, $\be$, and $T$.

\subsection{Three-body force}

In Refs.~\cite{tbf1,tbf2,tbf3,tbf4},
the TBF is constructed within the meson-exchange-current approach,
and we refer to these references for all lengthy technical details.
In this model,
the contributions due to two-meson exchanges
($\pi\pi$, $\pi\rho$, $\rho\rho$, $\sigma\sigma$, $\sigma\om$, $\om\om$),
involving Delta and Roper resonance excitation
and the important Z-diagram ($N\bar{N}$ excitation)
are included.
All parameters of the TBF model, i.e.,
the coupling constants and form factors,
are consistently determined to reproduce
the Argonne $V_{18}$ $NN$ interaction and the values can be found in
Refs.~\cite{tbf2,tbf3}.
Finally the TBF can be reduced to an
equivalent effective two-body force $V_3^{\rm eff}$ via a suitable
integration over the degrees of freedom of the third nucleon.
This procedure can be extended to finite temperature \cite{bbg4},
and the effective interaction $V_3^{\rm eff}(T)$ in $r$ space reads
\bal
 & \langle \rv_1',\rv_2' | V_3^{\rm eff}(T) | \rv_1,\rv_2 \rangle
 = \frac14 \text{Tr}\sum_n f(\kv_n,T) \int d\rv_3 d\rv_3'
\\[-1mm]&\times
 \phi_n^*(\rv_3') W_3(\rv_1',\rv_2',\rv_3'|\rv_1,\rv_2,\rv_3) \phi_n(\rv_3)
\nonumber\\&\times
 [1-\eta(r_{13}',T)] [1-\eta(r_{23}',T)]
 [1-\eta(r_{13},T)] [1-\eta(r_{23},T)] \:,
\nonumber
\eal
where $\phi_n$ is the wave function of the single nucleon in free space
and the trace is taken with respect to spin and isospin of the third nucleon.
The defect function $\eta(r,T)$ is directly related
to the temperature-dependent $G$ matrix.
$W_3(\rv_1',\rv_2',\rv_3'|\rv_1,\rv_2,\rv_3)$ represents the TBF,
which is given in detail in Ref.~\cite{tbf3}.
The result is an effective interaction with the operator structure
\bal
 V_3^{\rm eff}(\rv) &=
 V_I(r) + \ss V_S(r) + \tt\ss V_C(r)
\nonumber\\&
 +\, S_{12}(\hat{\rv}) \big[ \tt V_T(r) + V_Q(r) \big] \:,
\label{e:V3}
\eal
where $S_{12}(\hat{\rv}) =
3(\sv_1 \cdot \hat{\rv})(\sv_2 \cdot \hat{\rv}) - \sv_1 \cdot \sv_2$
is the tensor operator
and the components
$V_O,\; O=I,S,C,T,Q$
depend on the nucleon densities $\rho_{n,p}$ and temperature.
They are added to the bare potential $V_{NN}$ in the
Bethe-Goldstone equation for the $G$ matrix.

Note that the method of using an effective $NN$ interaction to treat the TBF
is an approximation that neglects certain many-body contributions \cite{Dyh16}.
The averaging procedure avoids the difficult problem of solving
the relevant Faddeev equation involving TBF.
It allows to include the direct and some single-exchange TBF diagrams in the
ladder summation of the BHF approximation,
but neglects in particular
the double-exchange TBF diagrams
\cite{def,tbf1,coonpi,coonrho}.
The individual sizes of these missing contributions have been estimated
to be of the order of 20\% \cite{coonpi}.
This approximation has been extensively used
and considered reliable in the past.
Going beyond it will require
a consistent inclusion of TBF into the hole-line expansion,
a considerable effort which might be achieved in the future.

\subsection{Treatment of total momentum}

Using the total and relative momentum,
\beq
 \Kv=\kv_1+\kv_2 \ ,\quad
 \kv=\frac12(\kv_1-\kv_2) \:,
\eeq
the BG equation~(\ref{e:bge}) can be transformed into
\bal
 &\delta_{\Kv\Kv'} \langle\kv|G_{\tau\tau'}(\Kv,\om,T)|\kv'\rangle
 = \delta_{\Kv\Kv'} \langle\kv|V_{\tau\tau'}(T)|\kv'\rangle
\nonumber\\
 &+ \sum_{\Kv''\kv''}\delta_{\Kv\Kv''}
 \langle\kv|V_{\tau\tau'}(T)|\kv''\rangle
 \frac{Q_{\tau\tau'}\!(\Kv'',\kv'',T)}{\om-e_\tau(\frac12\Kv''+\kv'')
 -e_{\tau'}(\frac12\Kv''-\kv'')}
\nonumber\\&\hskip6mm\times
 \delta_{\Kv''\Kv'} \langle\kv''|G_{\tau\tau'}(\Kv'',\om,T)|\kv'\rangle \:.
\eal
Generally, the nucleon interaction $V$ is independent of the total momentum.
However, the Pauli operator and the energy denominator depend on it.
Therefore, the BG equation can be written as
\bal
 &\langle\kv|G_{\tau\tau'}(\Kv,\om,T)|\kv'\rangle
 = \langle\kv|V_{\tau\tau'}(T)|\kv'\rangle
\label{e:gav}
\\
 &\hskip3mm+ \sum_{\kv''}
 \frac{
 \langle\kv|V_{\tau\tau'}(T)|\kv''\rangle\;
 Q_{\tau\tau'}\!(\Kv'',\kv'',T)\;
 \langle \kv''|G_{\tau\tau'}(\Kv'',\om,T)|\kv'\rangle
}
 {\om-e_\tau(\frac12\Kv''+\kv'')-e_{\tau'}(\frac12\Kv''-\kv'')}
\:.
\nonumber
\eal
For any given density, isospin asymmetry, and temperature,
the calculations of the s.p.~potential, Eq.~(\ref{e:u}),
need the full information of $G$ at arbitrary values of $\Kv$ and $\om$.
One therefore solves the BG Eq.~(\ref{e:gav}) on a $N_K \times N_\om$ grid,
where $N_K$ ($N_\om$) is the number of the $K=|\Kv|$ ($\om$) points.
Note that the value of the $G$ matrix should be independent
of the orientation of $\Kv$.

Such calculations were challenging several decades ago.
Also, since the value of the $G$ matrix is regarded to be insensitive
to the value of the total momentum $K$,
in the initial calculations of Brueckner theory \cite{bbgi},
an average-c.m.-momentum approximation was used
and the total momentum was approximated by the value
\beq
 \langle K_{\tau\tau'}^2 \rangle(k) =
 \frac{ \int_0^{k_F^{\tau}}\!\!d\kv_1 \int_0^{k_F^{\tau'}}\!\!d\kv_2
        \delta(k-\frac12|\kv_1-\kv_2|)\, (\kv_1+\kv_2)^2 }
 { \int_0^{k_F^{\tau}}\!\!d\kv_1 \int_0^{k_F^{\tau'}}\!\!d\kv_2
   \delta(k-\frac12|\kv_1-\kv_2|)} 
\eeq
at zero temperature.
This approximation has been widely adopted in former calculations
\cite{bbg1,bbg2,bbg3,bbg4,tbf2,df1}.
However, in the recent works of both BHF \cite{Nic06,Bal14,Tar16,For18,Lu19,Wei20}
and DBHF approaches \cite{th1},
the exact treatment of the total momentum has been used
and we thus also follow this way in the present calculations
to obtain more accurate results of the effective masses.

\section{Results}

\subsection{Equation of state}

\begin{figure}[t]
\vspace{-12mm}\hspace{-0mm}
\centerline{\includegraphics[scale=0.45]{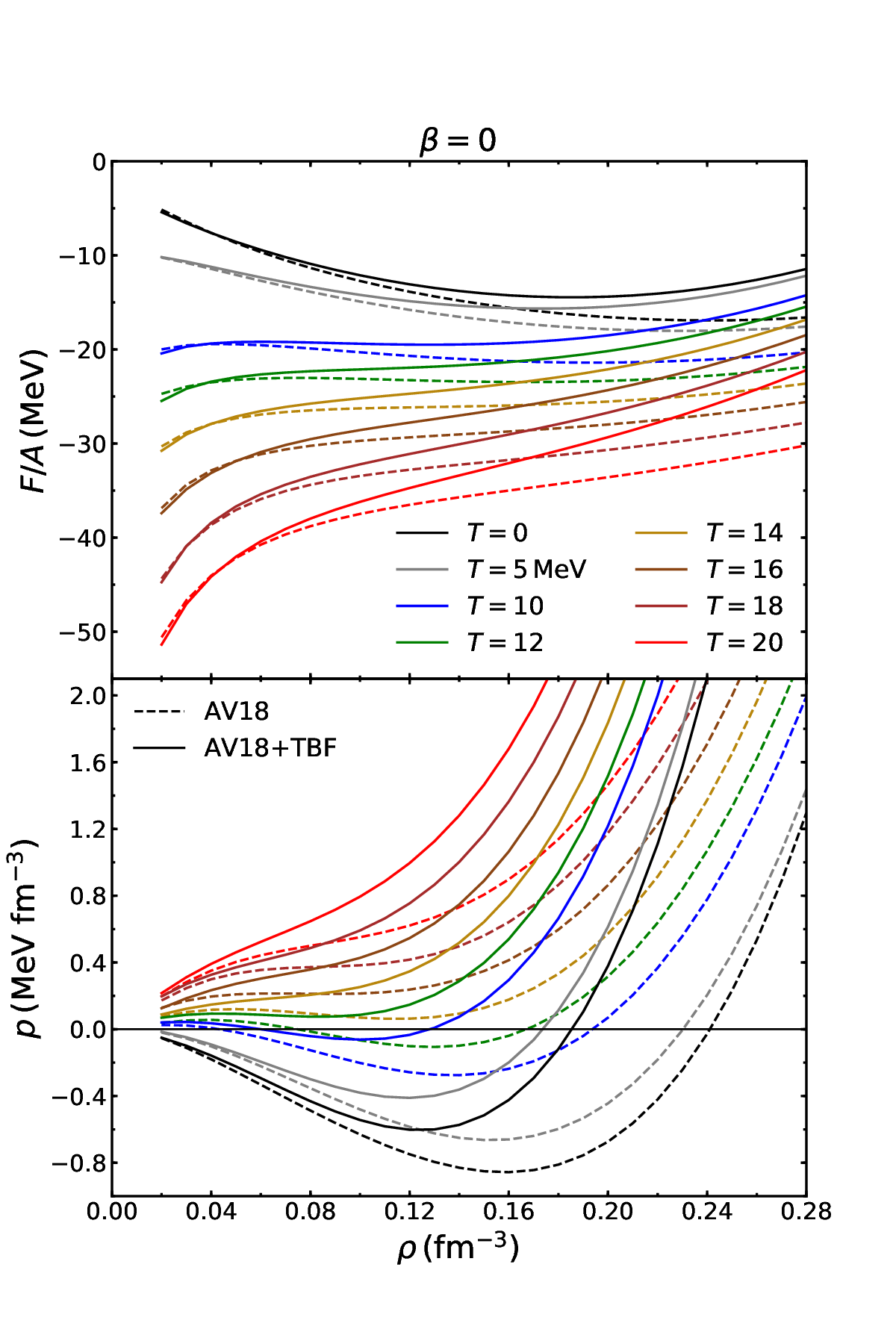}}
\vskip-10mm
\caption{
Free energy per nucleon (upper panel)
and pressure (lower panel)
of symmetric nuclear matter ($\be=0$)
as function of density
at $T$ = 0, 5, 10, 12, 14, 16, 18, and 20 MeV.
The solid and dashed curves are the results of
including or not TBF, respectively.
}
\label{f:eos}
\end{figure}

We first briefly discuss some aspects of the finite-temperature EOS
in our approach.
The zero- and finite-temperature V18 BHF EOS has been discussed in great
detail in several previous publications \cite{bbg4,Li10,Lu19,Wei20},
to which we refer for further information.
Here we only review some essential features:

The total energy density $\eps$ can be calculated from the $G$ matrix,
and the total entropy density $s$ can be evaluated in the approximation
of a noninteracting Fermi gas of quasiparticles
in the mean field $U_\tau(\kv)$ \cite{Lej86,Bom94}.
Then the free energy density $f=\eps-Ts$,
the chemical potentials $\mu_i=\partial f/\partial\rho_i$,
and the pressure $p=\rho^2 d(f/\rho)/d\rho$
can be computed according to the standard thermodynamic relations.
The obtained finite-temperature EOS of symmetric nuclear matter
(free energy per nucleon $F/A=f/\rho$ and pressure $p$)
is reported in Fig.~\ref{f:eos}.
The important role of TBF,
which act increasingly repulsive with density
and correct the nuclear saturation point of cold matter,
is clearly reckognized.
With the inclusion of TBF, the resulting saturation density is $0.186\fm3$
and the energy per baryon at saturation is $-14.5\mev$.
They are somewhat different from the values
($0.198\fm3$, $-15.0\mev$) reported in the original paper \cite{tbf1,tbf2},
indicating the effects caused by the exact treatment of the c.m.~momentum.
Regarding finite temperature,
similar critical temperatures
for the liquid-gas phase transition as in previous calculations \cite{bbg4}
are predicted:
about $13\mev$ ($16\mev$) with (without) the inclusion of TBF.
A similar decrease of the critical temperature due to TBF
is obtained using different $NN$ interactions
or other microscopic approaches \cite{Car18}.

\begin{figure*}[t]
\centering
\vskip-6mm\
\centerline{\includegraphics[scale=0.49,clip]{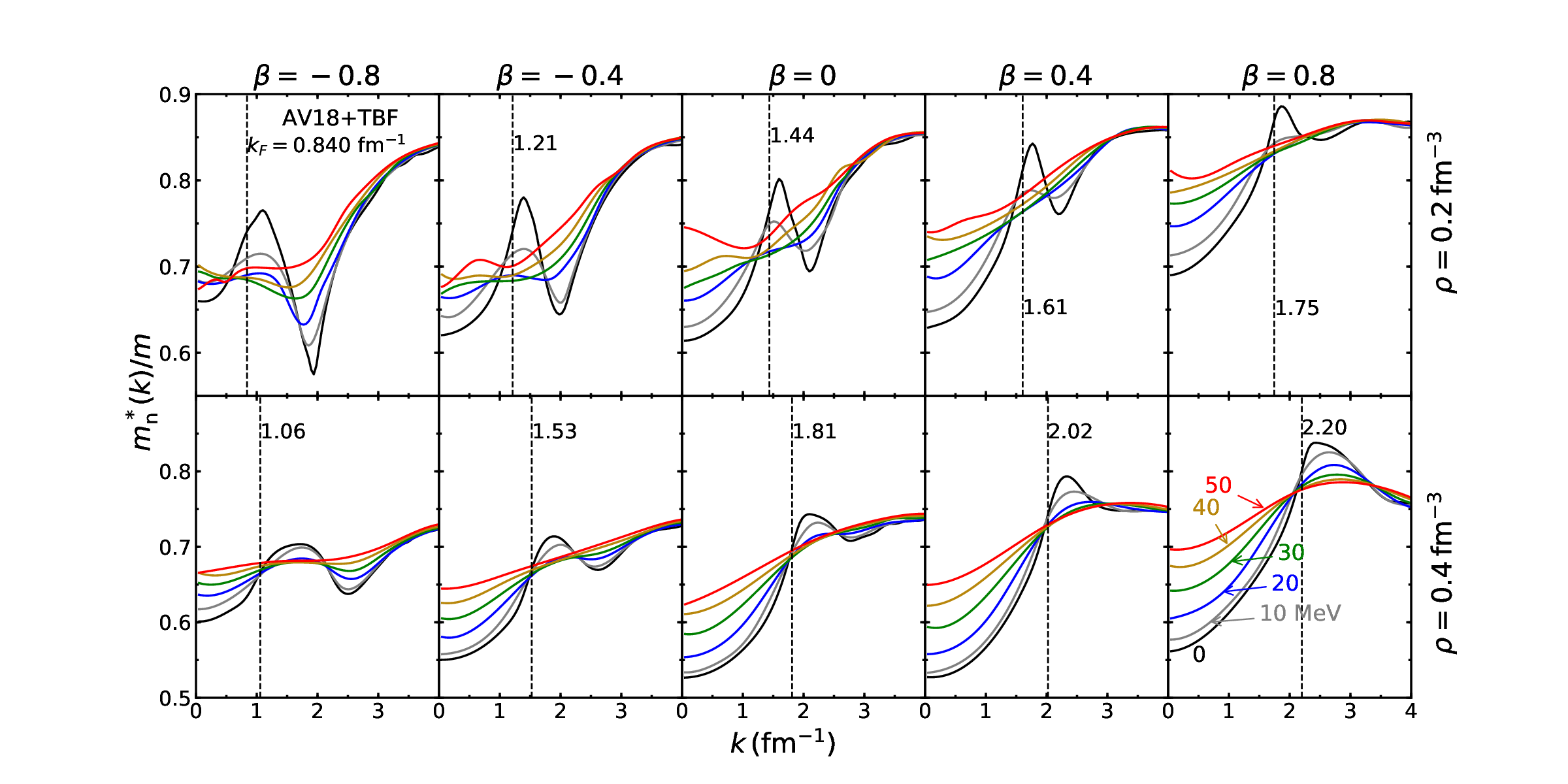}}
\vskip-4mm
\caption{
Neutron effective mass as a function of momentum $m_n^*(k)/m$,
Eq.~(\ref{e:ms}),
at temperatures $T = 0,10,20,30,40,50\mev$,
densities $\rho = 0.2,0.4\fm3$,
and asymmetries $\be=0,\pm0.4,\pm0.8$.
The adopted nucleon force is the Argonne $V_{18}$ potential
plus the microscopic TBF.
The vertical dashed lines indicate the neutron Fermi momenta.
For the proton one has $m_p^*(\be)=m_n^*(-\be)$.
}
\label{f:sp}
\end{figure*}

\begin{figure*}[t]
\centering
\vspace{-3mm}
\includegraphics[scale=0.45]{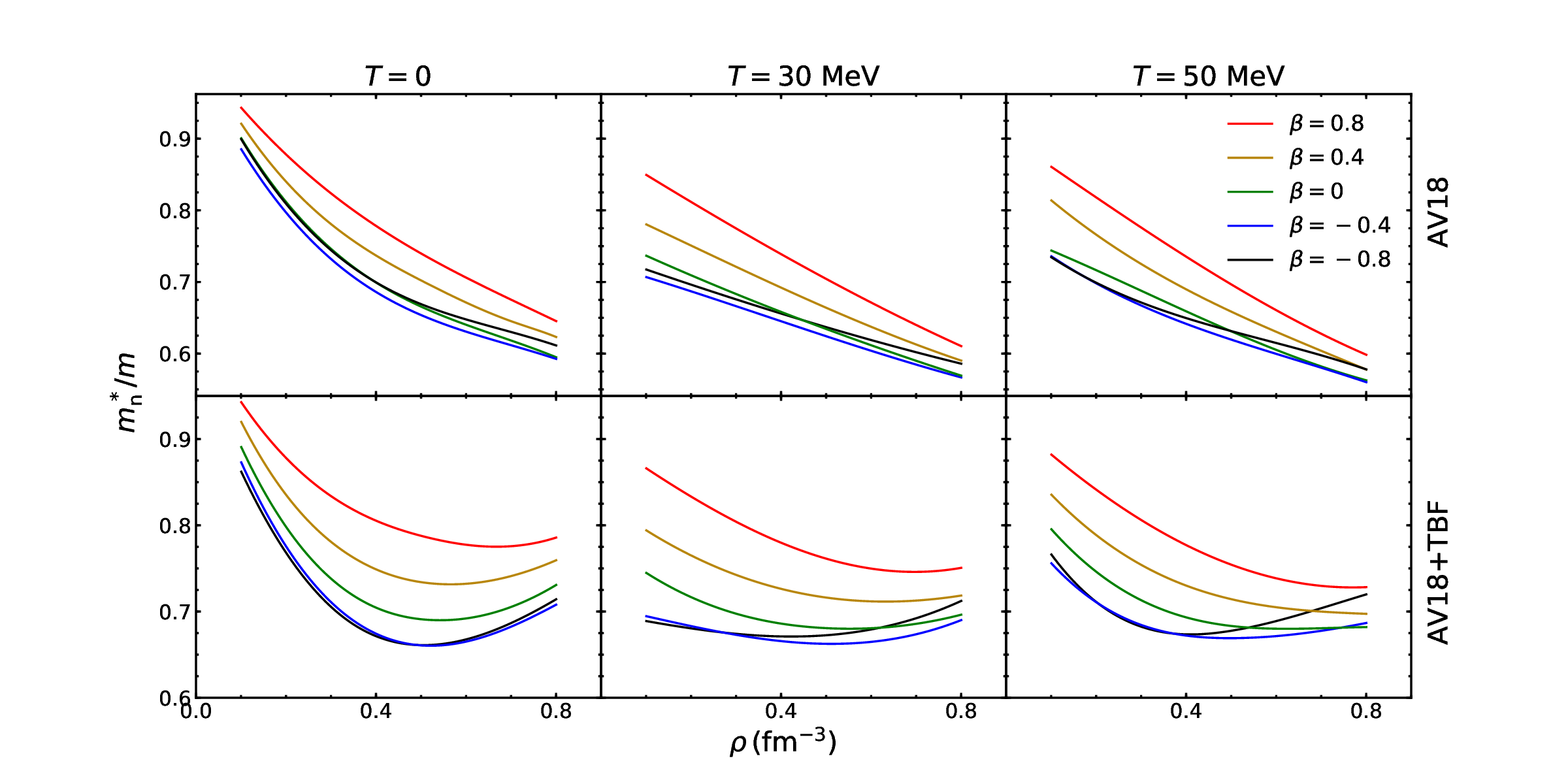}
\vskip-5mm
\caption{
Neutron effective mass as a function of density,
for temperatures $T = 0,30,50\mev$
and asymmetries $\be=0,\pm 0.4, \pm 0.8$,
with and without the TBF contribution.
For the proton one has $m_p^*(\be)=m_n^*(-\be)$.}
\label{f:tbf}
\end{figure*}

\begin{figure*}[t]
\centering
\vspace{-6mm}
\includegraphics[scale=0.45]{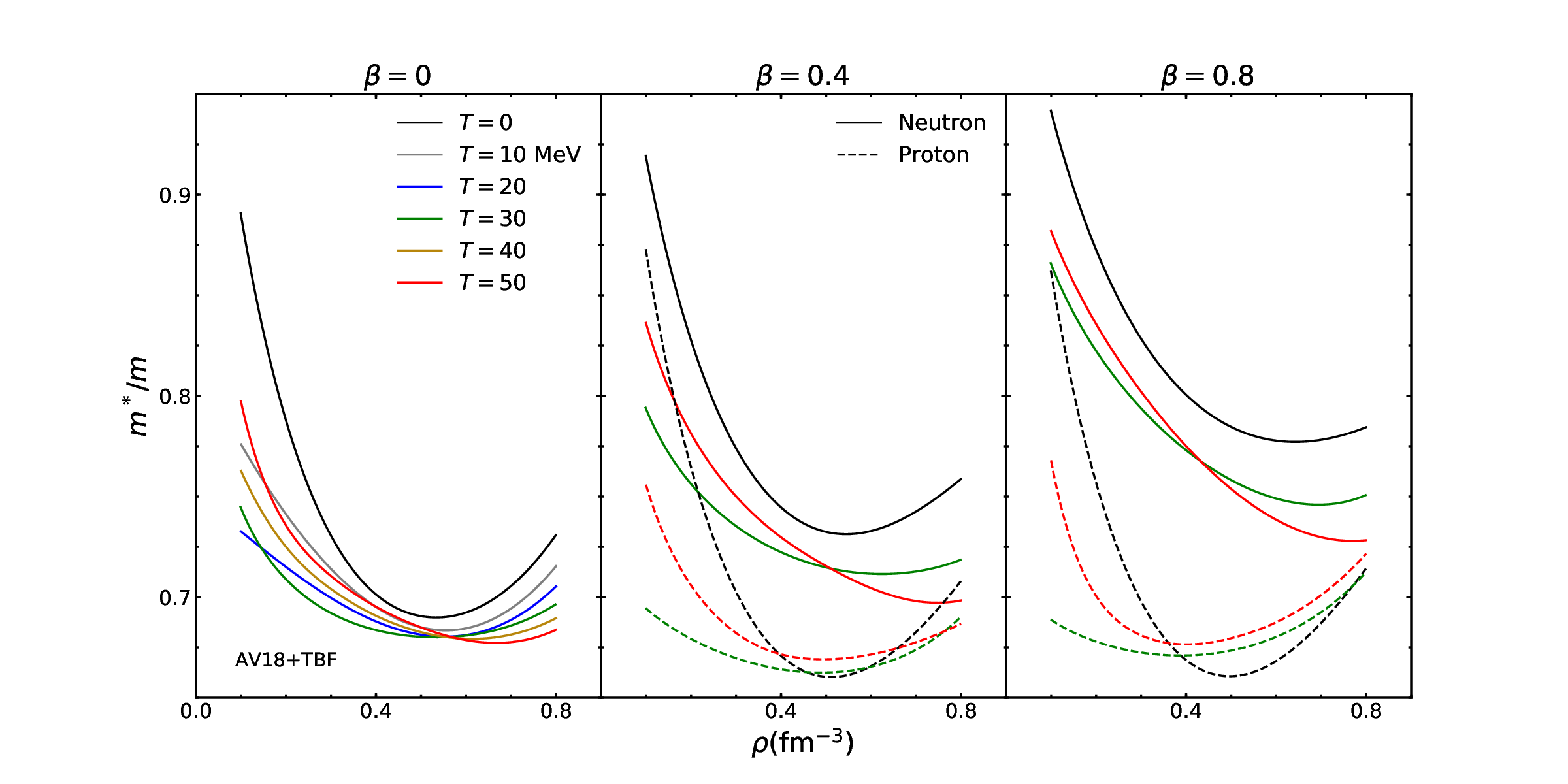}
\vskip-4mm
\caption{
Neutron/proton effective mass as a function of density
at different temperatures $T = 0,10,20,30,40,50\mev$ and
asymmetries $\be=0,0.4,0.8$.
The calculations are done including the TBF contribution.
}
\label{f:beta}
\end{figure*}

\begin{figure}[t]
\vspace{-8mm}\hspace{-0mm}
\centerline{\includegraphics[scale=0.45]{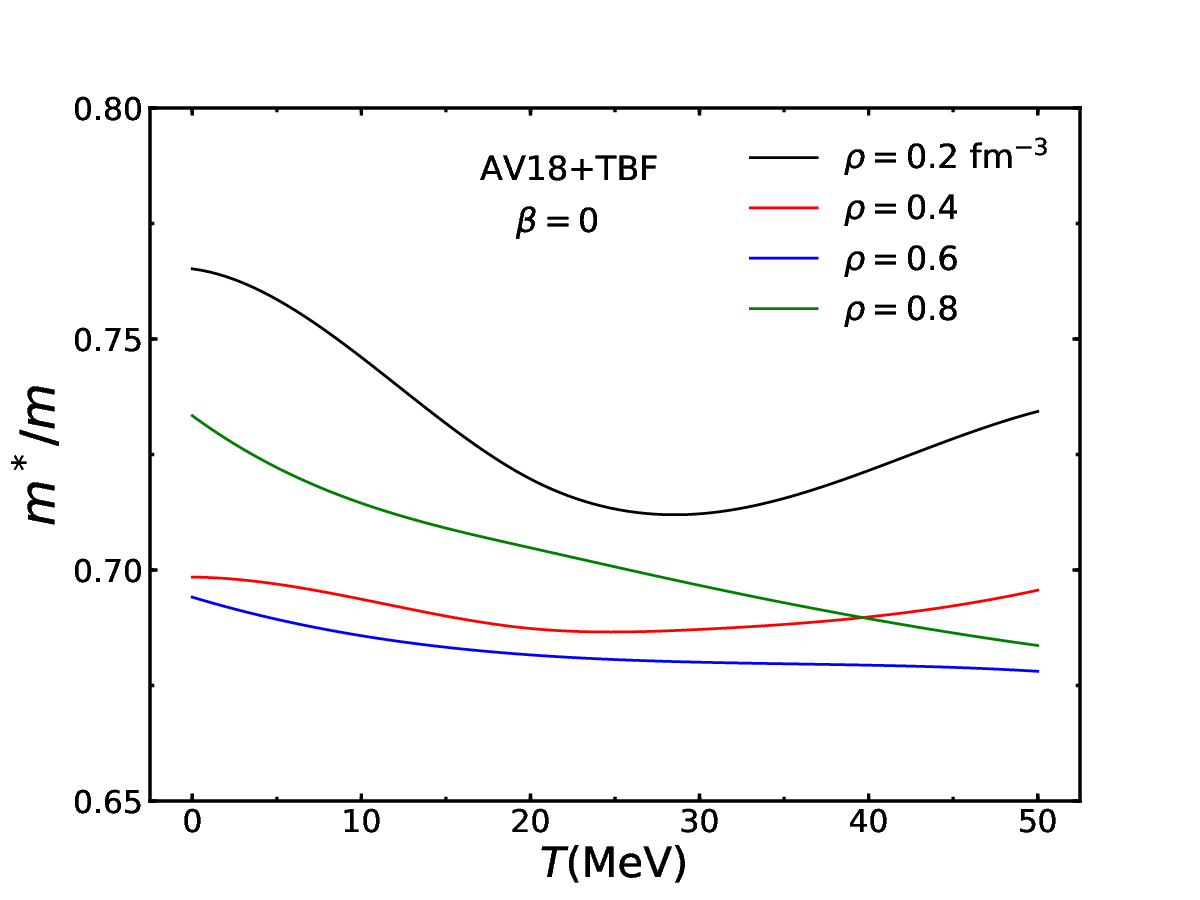}}
\vskip-3mm
\caption{
Nucleon effective mass of symmetric nuclear matter ($\be=0$)
with the inclusion of TBF
as a function of temperature
for fixed densities $\rho= 0.2,0.4,0.6,0.8\fm3$.
}
\label{f:snm}
\end{figure}

\subsection{Effective masses} \label{s:effmass}

We then show in Fig.~\ref{f:sp} the momentum dependence
of the neutron effective mass
at various temperatures $T = 0,10,20,30,40,50\mev$,
densities $\rho = 0.2,0.4\fm3$,
and isospin asymmetries $\be=0,\pm0.4,\pm0.8$.
Due to isospin symmetry, the proton and neutron effective masses are related by
$m^*_p(\be)=m^*_n(-\be)$.
The zero-temperature Fermi momenta
$k_F^{n/p} = [3\pi^2(1\pm\be)\rho/2]^{1/3}$
are shown by vertical lines.
The temperature effects are generally more significant at low momentum
and most evident around $k_F$,
where higher temperatures flatten the curves.
This is related directly to the smoothing of the sharp Fermi surface
and consequently of the s.p.~potential around the Fermi momentum,
and is a general feature for different choices of the $NN$ potential and TBF
\cite{bbg4,Lej86,Bal88,Bom94,Bom06,Zuo06}.

In the following Figs.~\ref{f:tbf},\ref{f:beta},\ref{f:snm}
we present the detailed results for the effective mass
$m^* \equiv m^*(k_F)$
spanning the whole asymmetry range
in a density domain up to $0.8\fm3$,
and a temperature up to $50\mev$.
The calculations are done with and without the TBF contribution.

One should mention in this context that
at low densities (below $\sim 0.1\fm3$)
the nuclear matter system can minimize its energy
by forming light clusters such as deuterons,
or particularly strongly bound alpha particles \cite{Typ10}.
In theoretical calculations,
such as the extended BHF approach,
the in-medium $T$-matrix method,
and the self-consistent Green's function method,
the effective interaction including all the ladder-diagram contributions
always encounters a singularity leading to unstable results at low densities
\cite{Eme59,Dic88,Von91,Are15},
which is related to the onset of formation of the deuteron bound state.
Moreover, since the $NN$ interaction models are fixed at low energy scales,
the point-particle picture also becomes unrealistic
at high densities (above $\sim 0.8\fm3$),
where quark degrees of freedom should be considered.
Therefore both at low and high densities
the BHF calculation
should be taken with caution.

In Fig.~\ref{f:tbf} we compare the density dependence of the effective mass
with or without TBF, at different temperatures and asymmetries.
As already mentioned in the introduction
and shown in Fig.~\ref{f:eos},
the inclusion of TBF is important for reproducing the saturation properties
of nuclear matter.
We see here that it also changes the behavior of $m^*(\rho)$ at high density:
After the inclusion of TBF,
$m^*$ rises with density after reaching a certain minimum at $\rho_{\rm min}$,
as already observed in the works of Ref.~\cite{Bal14,Li16} at zero temperature.
This results from the repulsive nature of the TBF \cite{tbf1,tbf2,tbf3}
and resembles the DBHF result \cite{Dal05}.
The general effect of temperature is to smooth out the rising
of the effective mass caused by the TBF contribution,
shifting $\rho_{\rm min}$ to higher values.
Isospin asymmetry causes the minority component to acquire a lower
effective mass than the isospin partner.

To see more clearly the interplay between temperature effect
and the TBF contribution,
we show in Fig.~\ref{f:beta} a comparison at different temperatures
and asymmetries for both neutron and proton effective masses.
We see again the flattening effect of temperature at high density.
At low density the temperature will first reduce
(removal of the s.p.~`wiggle')
and then increase the effective mass, see Fig.~\ref{f:sp}.
This is the case for both neutron and proton and different asymmetries.

Fig.~\ref{f:snm} is devoted to the comparison of the density and temperature
dependence of the effective mass.
We present the results with TBF and for symmetric nuclear matter.
The curves are plotted for a set of densities ranging from 0.2 to $0.8\fm3$,
and temperature from 0 to $50\mev$.
Comparing with the left panel of Fig.~\ref{f:beta},
one concludes that
the effective mass is generally more sensitive to density than to temperature.
The temperature dependence tends to be pronounced at low density
and the density dependence tends to be pronounced at low temperature.
The behavior of $m^*$ with increasing density is very similar
at different temperatures:
$m^*$ first decreases and then increases with density.
This is mainly due to the increasingly dominating role of the TBF,
which has a repulsive nature.
The behavior of $m^*$ with increasing temperature is, however,
not straightforward for different densities.
Due to the competitive effect between the density and the temperature,
at intermediate densities such as $\rho=0.4,0.6\fm3$,
the temperature dependence is very limited.
At low density such as $\rho=0.2\fm3$,
$m^*$ first decreases and then increases with temperature,
as also observed in Fig.~\ref{f:beta}.
At high density such as $\rho=0.8\fm3$,
the flattening effect of temperature dominates
and $m^*$ decreases monotonically with temperature.

We conclude this section by commenting that
compared to our BHF results,
different many-body approximations may predict
somewhat different effective masses,
resulting from changes in the interaction models and/or the many-body frameworks,
but the qualitative results are usually similar
\cite{Hub98,Rio05}. 
It appears that an overall larger pressure yields a larger effective mass,
which reduces the increase with temperature of the free energy
and therefore leads in turn to a lower critical temperature.
In particular, recent investigations employing chiral $NN$ and $NNN$ forces
\cite{Car18,Car19n,Car19s}
require much stronger TBF in order to compensate the too strong attraction
of the soft-core chiral potentials.

\begin{table*}
\squeezetable
\caption{
Fit parameters of the neutron effective masses
in the functional form of Eq.~(13).}
\renewcommand\arraystretch{1.5}
\begin{ruledtabular}
\begin{tabular*}{\hsize}{@{}@{\extracolsep{\fill}}lcccccccccccccccccccc@{}}
      & $a_1$ &$b_1$ &$c_1$ &$a_2$ &$b_2$ &$c_2$ &$a_3$ &$b_3$ &$c_3$
      & $a_4$ &$b_4$ &$c_4$ &$d_1$ &$d_2$ &$d_3$ &$d_4$ &$d_5$ &$d_6$\\
\hline
V18   & 0.607 &-0.070 & 0.0687&-0.037 & 0.0477&-0.0156& 0.256 &-0.0797& 0.0177
      &-0.372 & 0.157 &-0.0590& 0.0051&-0.126 & 0.253 &-0.097 &-0.0273& 0.447\\
+TBF  & 0.102 &-0.094 & 0.0680& 0.699 & 0.0354& 0.0133& 0.750 &-0.0662& 0.0349
      &-0.941 & 0.235 &-0.0747&-0.0135&-0.181 & 0.421 &-0.367 &-0.0150& 1.010\\
\end{tabular*}
\end{ruledtabular}
\label{t:fit}
\end{table*}

\begin{figure}
\vspace{-18mm}\hspace{-0mm}
\centerline{\includegraphics[scale=0.42,clip=no]{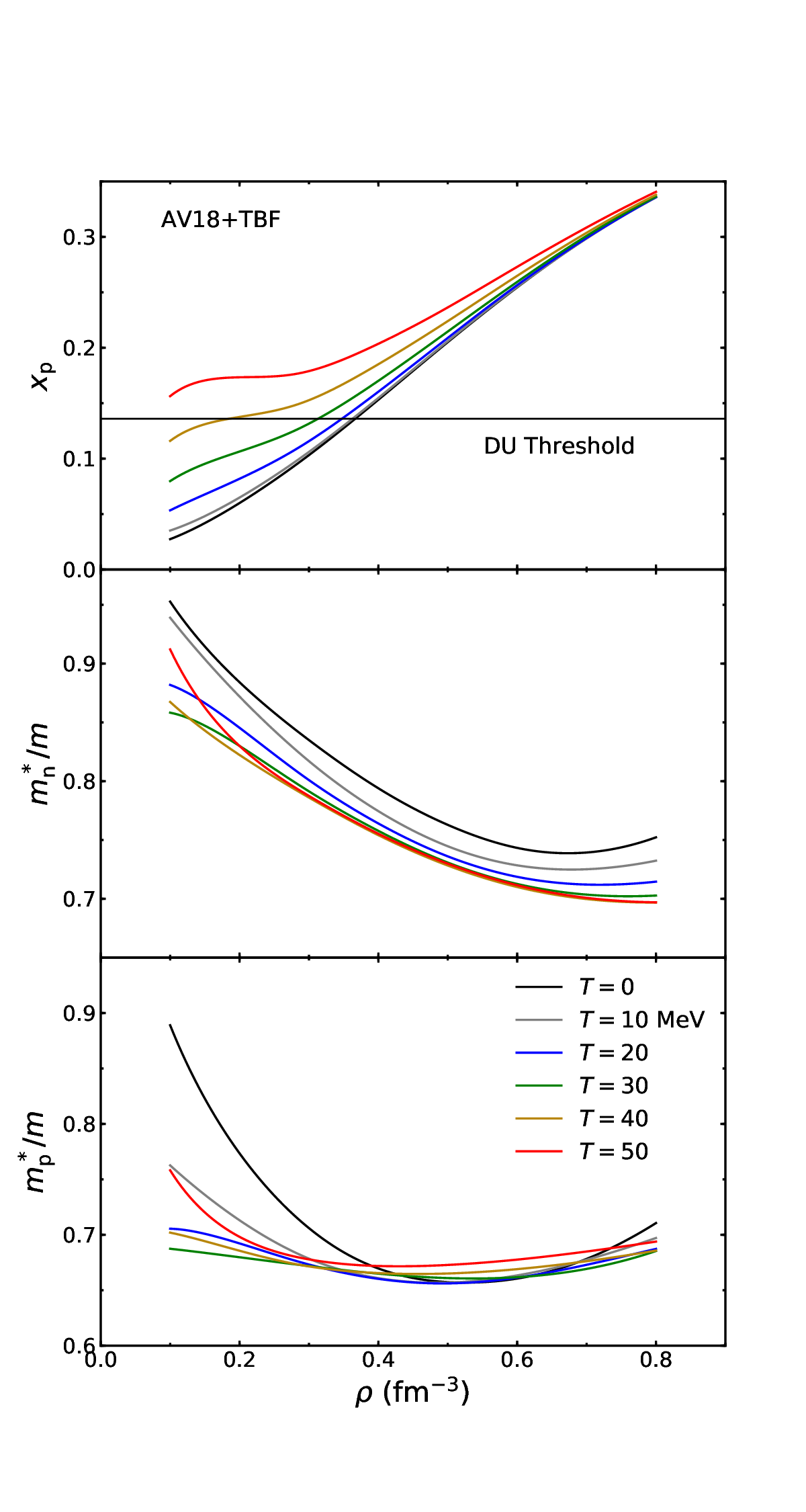}}
\vskip-13mm
\caption{
Proton fraction (upper panel)
and neutron/proton effective masses (middle and lower panels)
for betastable NS matter as functions of density at different temperatures
$T=0,10,20,30,40,50\mev$.
The calculations are done with the TBF contribution.
The horizontal line indicates the threshold proton fraction
of the direct Urca neutrino emission process.
}
\label{f:ns}
\end{figure}

\subsection{Fit formula}

One of the main goals of the present study is to provide
easy-to-use microscopic nuclear input for various astrophysical systems.
We therefore fit the numerical results of the effective mass
by an analytic representation
(with the three independent variables
density $\rho$, asymmetry $\be$, and temperature $T$),
extending the zero-temperature formulas \cite{Bal14}.
We choose the following empirical form:
\bal
 \frac{m^*_n}{m}(\rho,\be,T) &=
    a_1 + b_1\be + c_1\be^2 + (a_2 + b_2\be + c_2\be^2)\rho
\\\nonumber
 &+ \big[a_3 + b_3\be + c_3\be^2 + (a_4 + b_4\be + c_4\be^2)\rho\big]t
\\\nonumber
 &+ (d_1 + d_2t + d_3t^2)/\rho+(d_4+d_5t+d_6t^2)\ln\rho \:,
\eal
valid for the domain
$0.1\fm3 \le \rho \le 0.8\fm3$,
$-1 \le \be \le 1$,
and $10\mev \le T \le 50\mev$,
where $t=T/(100\mev)$ and $\rho$ is given in $\!\!\fm3$.
The parameters of the fit are listed in Table~\ref{t:fit}
with and without TBF.
The standard deviations are 0.010/0.008, respectively.
The results for protons are obtained as $m^*_p(\be)=m^*_n(-\be)$.
We remark that these fits should only be employed in the regime of homogeneous
nuclear matter modeled by BHF theory, $\rho\gtrsim0.1\fm3$.
Lower densities are characterized by the appearance of cluster structures,
where other theoretical approaches must be used,
see also the comments in Sec.~\ref{s:effmass}.

\subsection{Betastable matter}

Finally we report in Fig.~\ref{f:ns}
the calculations of hot beta-stable NS matter
at different temperatures.
The proton fraction $x_p$ and the effective masses $m^*_{n,p}$
are plotted as functions of density,
for temperatures from 0 to $50\mev$.

The direct Urca (DU) process,
corresponding to neutron $\beta$-decay and its inverse reaction
$ n \rightarrow p + e + \bar{\nu}_e,\ p + e \rightarrow n + \nu_e $,
is the most efficient neutrino cooling process \cite{gle}.
It only occurs in cold NSs if the proton fraction is larger than a
critical threshold
($x_p\approx0.138$, slightly dependent on the muon fraction),
such that energy and momentum can both be conserved
at sufficiently high density for these two persistent reactions.

We see in the upper panel that the BHF EOS with V18+TBF
predicts a relatively low threshold density for the DU process,
close to $\rho_{\rm DU}\approx0.38\fm3$ at zero temperature,
ensuring fast cooling being active in nearly all NSs
(see the discussion in Refs.~\cite{Tar16,For18,Wei19}, for example).
Finite temperature increases the proton fraction
due to the presence of thermal leptons
and therefore decreases $\rho_{\rm DU}$.
The temperature mainly affects the low-density domain of the proton fraction,
as already observed in our previous works
\cite{BS10,Li10,BSL11,Che12,Li13,Li15}.
In the middle panel, we see that the direct and indirect
(decrease of the neutron partial density)
effects of increasing temperature lead to a decrease of the
neutron effective mass at nearly all densities.
The values are somewhat higher than in symmetric matter, see Fig.~\ref{f:beta}.
The proton effective mass, displayed in the bottom panel,
shows a similar flattening behavior with increasing temperature,
with a value of about 0.7 and a
weak dependence on the temperature and density
for $\rho \gtrsim 0.4\fm3$.

\section{Summary}

The nucleon effective mass at finite temperature
is of fundamental importance for nuclear astrophysics,
but an evaluation of the s.p.~properties is usually not easy and model dependent.
Previous works on the temperature dependence of the effective mass
showed nontrivial behavior for the required ranges
of nucleon density and isospin asymmetry
in dynamical astrophysical systems of interest.
So we performed the calculation of $m^*(\rho,\be,T)$
from realistic nucleon forces within a microscopic model.
We used the BHF method extended to asymmetric nuclear matter
and finite temperature,
employing the realistic Argonne $V_{18}$ force together with
consistent microscopic TBF.

We studied the interplay of the $\rho, \be, T$ dependence of the effective mass
with and without the TBF contribution.
Finite temperature in general lowers the effective mass,
in particular at low and high densities.
TBF increase the effective mass at high density due to their repulsive character,
but finite temperature weakens this effect.
Altogether, the temperature dependence is modest in comparison
to the density dependence,
but the specific behavior can be different in different density domains.

The dependence $m^*(\rho,\be,T)$
has been accurately parametrized by a carefully chosen analytical formula,
to be conveniently used for the study of NS cooling,
merger simulations, core collapse supernovae, heavy-ion collisions, etc.
We have also discussed the temperature dependence of the proton fraction
and the nucleon effective mass in betastable NS matter,
and the influence on the DU process in a hot star.
The present results might be used for the study of the thermal evolution
of a PNS or a NS merger event,
which we will explore in a future work.

\medskip
\acknowledgments

We would like to thank J.~M.~Dong, Z.~H.~Li, and W.~Zuo for valuable discussions.
We appreciate great help of Li Xue for computation on the XMU-astro clusters.
The work was supported by the National Natural Science Foundation of China
(Nos.~11873040, 11505241, 11775276),
and the Youth Innovation Promotion Association
of the Chinese Academy of Sciences.
We further acknowledge partial support from ``PHAROS,'' COST Action CA16214.


\end{document}